\documentclass[12pt]{revtex4}
\usepackage{epsfig}

\def\bx{{\bf x}}
\def\bv{{\bf v}}
\def\br{{\bf r}}

\def\bp{{\bf p}}
\def\bn{{\bf n}}
\def\bl{{\bf L}}
\def\bs{{\bf S}}
\def\bD{{\bf \Delta}}
\usepackage[mathscr]{euscript}
\begin{document}
\bibliographystyle{apsrev}

\title{Chaos and Order in Models of Black Hole Pairs}

\author{Janna Levin}
\affiliation{Department of Physics and Astronomy, Barnard College,
Columbia University,
3009 Broadway, New York, NY 10027}

\begin{abstract}
Chaos in the orbits of black hole pairs has
by now been confirmed by several independent groups. 
While the chaotic behavior of binary black hole orbits is no longer
argued, it remains difficult to quantify the importance of chaos to
the evolutionary dynamics of a pair of
comparable mass black holes. None of our existing approximations are
robust enough to offer convincing quantitative conclusions in the most
highly nonlinear regime. 
It is intriguing to note that
in three different approximations to 
a black hole pair built of a spinning black hole and a {\it
  non-spinning} companion,  
two approximations exhibit chaos and one approximation does not. The
fully relativistic scenario of 
a spinning test-mass around a Schwarzschild black hole shows chaos, as does the
Post-Newtonian Lagrangian approximation. However, the approximately
equivalent Post-Newtonian Hamiltonian approximation does not show
chaos when only one body spins.
It is well known in dynamical
systems theory that one system can be regular while an approximately
related system is chaotic, so there is no formal conflict. However,the 
physical question remains, Is there chaos for comparable mass
binaries when only one object spins? We are unable to answer this
question given the poor convergence of the Post-Newtonian
approximation to the fully relativistic system. A resolution awaits
better approximations that can be trusted in the highly nonlinear regime.
\end{abstract}


\maketitle

\vfill\eject
 
\setcounter{section}{1}

Isolated black holes are beautifully simple, as are the orbits of test
particles around them.
The elegance of the Kerr metric for a rotating
black hole is impressive. Impressive too is Carter's \cite{carter}
characterization of the orbits 
of a test-particle in the Kerr spacetime.
Carter \cite{carter} found enough constants of
motion to prove that the geodesics around a Kerr black hole were
perfectly regular; that is to say, integrable. However, general
relativity no longer looks simple once we begin
to consider two black holes. Two black holes pose a notoriously
difficult problem and as yet have defied all attempts at a
solution. What's more, the dynamics of two spinning black holes even
shows chaotic episodes
\cite{{me},{melong},{HB},{maeda1}}.

In retrospect, it is no surprise that there is chaos in the orbits of
two spinning black holes. The inner orbits around 
a non-spinning Schwarzschild black hole, although absolutely
integrable, already
provide prime terraine for the onset of chaos. Most notable is the
presence
of a hyperbolic fixed point, better known as an unstable circular
orbit, with which is associated a homoclinic orbit - an orbit that
approaches the circular orbit both in the infinite past and the
infinite future. It is well known, that under perturbation
homoclinic orbits can give
rise to a homoclinic tangle, an infinite intersection of the stable
and unstable manifolds of a hyperbolic fixed point. The presence of a
spinning companion can give rise to a homoclinic tangle and in the
increasingly complicated set of bound orbits that results one finds
chaos. 

Although, in retrospect it is no surprise, the existence of
chaos in the orbits of two spinning black holes has met with intense
resistance. Still, there it is. Chaos was originally found and
confirmed in several different
approximations to the two-body problem
\cite{{me},{melong},{HB},{maeda1},{menjc_comment},{menjc},{menjc2}}.
Whether or not chaos will affect future detections from the
gravitational wave observatories is still a cloudy issue but it is
fair to say the experiments will not be very adversely affected
and in an ideal world we might even have the opportunity one day to
witness the onset of chaos.

A fascinating subtlety which deserves a little attention is
the following: In three different approximations to the same physical
system, two exhibit chaos and one does not. The physical system is one
non-spinning black hole with a spinning companion. The three different
approximations are (1) the extreme-mass-ratio limit of a Schwarzschild
black hole orbited by a spinning companion (2) the Post-Newtonian
(PN) {\it Lagrangian} formulation of the two-black hole system
with one body spinning and (3) the
PN {\it Hamiltonian} formulation of the two-black hole system
with one body spinning. All three
systems are purely conservative, so radiation reaction is effectively
turned off. 

Saying this another way, there is chaos in the full relativistic
system when only one body spins. The
chaos is absent in the PN Hamiltonian approximation to this fully
relativistic system -- at least it is absent up to order 3PN
\cite{{gk},{gklong},{note}}. This is a 
reflection of the PN expansion's poor convergence to the full
nonlinearities of general relativity. The chaos
appears in an approximation to the approximation, namely the PN
Lagrangian formulation but at an order higher than the
approximation 
can be trusted. 

There is actually no explicit conflict between these results. They can
all be correct.
There can be
chaos in the full system that goes away in an approximation to the
system.
And the chaos that went away in an approximation can reappear in a
related system.
It is well-known in dynamical systems theory that a regular, that is to
say nonchaotic, system can become chaotic under a small perturbation. 
In fact there is a famous theorem that helped 
identify the locus of chaos under small perturbations, the
Kolmogorov-Arnol'd-Moser (KAM) theorem \cite{{kol},{ar},{mos}}.

A quick sketch of the argument begins with a regular
Hamiltonian system $H_0$ with $N$ coordinates and $N$
conjugate momenta. If there are $N$ constants of motion, then the
system is decidedly integrable and not at all chaotic. A canonical
transformation to action-angle coordinates, (${\bf \Theta},{\bf I})$,
can be performed so that each of the
new conjugate momenta are set equal to one of the $N$ constants of
motion and the Hamilton equations become
\begin{equation}
\dot {\bf I}=0 \ \ , \quad\quad \dot{\bf \Theta}=\frac{\partial H}{\partial {\bf I}}=
{\bf \omega}({\bf
  I})\ .
\end{equation}
The angular frequencies only depend on the constant ${\bf I}$ and so
are also constant. Therefore the motion in each coordinate direction
is cyclical and orbits are
confined to an $N$-dimensional
torus. Now the KAM theorem shows that under a small perturbation, the
motion will remain quasiperiodic for most initial data and that the
remaining orbits that do not remain quasiperiodic occupy a region of
phase space as small as the perturbation is small. 
Generally speaking, these latter orbits correspond to resonant tori
which are destroyed under perturbation and allow for the onset of chaos.
The gist is that one system can be regular while
an approximately related system is chaotic. So, for instance,
the Lagrangian formulation of the PN two-black hole dynamics can show
chaos while the approximately equivalent Hamiltonian formulation does not.
(It should be emphasized that the Lagrangian and Hamiltonian
equations of motion are related by a gauge transformation {\it and an
  approximation}.)

The deeper question that emerges is this: If there is chaos in one
formulation of the dynamics but not in an approximately equivalent
formulation, which one is right? That is, Is there chaos for one
spinning body or not? The answer is this:
The full relativistic system
exhibits chaos when only one body spins
\cite{maeda1}. That should be the final story.
The absence of chaos in the PN Hamiltonian
approximation to the case of one body spinning must be a consequence
of
the slow convergence of the PN expansion to the fully relativistic
system. Some of the nonlinearity is omitted in the approximation.
So physically, the fully relativistic result is the correct result.

However, it was also show in Ref.\ \cite{maeda1} that around a
Schwarzschild black hole a spinning test particle will become chaotic
only for unphysically large spins. If the heavy black hole has mass
$m_1$ and the light companion has mass $m_2$, then the spin of the
light companion must be $S_2 > 1 \mu M = 1 m_2^2(m_1/m_2)$, where
$M=m_1+m_2$ and $\mu=(m_1m_2)/M$.
This spin is much larger than maximal given the
extreme mass ratio $m_1/m_2 \gg 1$.
The physical question we can worry about now is, 
Will there be chaos for {\it comparable mass} binaries at a
physical value of the spin, $S\le m^2$, 
when only one object spins? This has yet to be
determined.

For comparable mass systems with physically accessible spins, 
the 2PN Hamiltonian dynamics says no, there is no chaos if only one
object spins, while the 2PN
Lagrangian dynamics says yes, there can be chaos even if only one body
spins. Neither is
definitive. Afterall, the 2PN Hamiltonian approach says there is no
chaos in the extreme mass ratio case when we know that there is chaos in the
fully relativistic system. On the other hand, the chaos seen in the
2PN 
Lagrangian system is of higher order than the approximation can be
trusted.

The physical question can only be resolved when the chaos appears or
disappears consistently at the same order as the approximation is
valid. And so I do not claim to resolve it here. Instead, I take 
a moment in the following section to give a quick demonstration of the
difference 
between the Hamiltonian and Lagrangian approaches and to show that the
former
is regular and the latter allows chaos when only one body spins.

\section{The PN Hamiltonian formulation}

The absence of chaos in the Hamiltonian formulation was recently
argued in Ref.\ \cite{gk} for
the dynamics of two compact objects 
when only one of the bodies spins. (They find similary that there can
be no chaos when the
binaries are of equal mass.) To be clear, the dynamics is conservative,
computed to second order in the Post-Newtonian expansion, and spin
effects
are limited to the spin-orbit couplings only. In all other situations,
there can be chaos - if both objects spin and are not of equal mass
and/or spin-spin couplings are included.
When (i) only
one body spins or (ii) when the binaries are equal mass,
the argument against chaos in the Hamiltonican dynamics 
is that there are enough (exactly conserved)
constants of motion to prove that the system
is integrable in these two simplified cases. And what's more, the
authors were able to find parametric solutions to the Hamiltonian dynamics
if only one body spins.
In accord with their claim, the technique of fractal basin boundaries used
in Ref.\ \cite{{me},{melong},{menjc}} indeed confirms that there is no chaos
in the Hamiltonian formulation when only one body spins, as will be
shown here.
We will confirm that the basin boundaries
for
this case are smooth and not fractal and therefore are entirely
consistent with regular, nonchaotic dynamics. 

The Hamiltonian is currently available to 3PN order. However, for
comparison with the Lagrangian formulation, we will only write the
Hamiltonian explicitly up to 2PN order. 
The reduced 2PN-Hamiltonian in ADM coordinates is (with ${\bf r}$
measured in units of $M$ and 
${\bf p}$ measured in units of $\mu$)
\begin{equation}
H=H_{N}+H_{1PN}+H_{2PN}
\end{equation}
with terms
\begin{equation}
H_N=\frac{{\bf p}^2}{2}-\frac{1}{r} \quad ,
\end{equation}
\begin{eqnarray}
H_{1PN}=\frac{1}{8}\left (3\eta-1\right ) \left ({\bf p}^2\right )^2
-\frac{1}{2}\left [\left (3+\eta\right ){\bf p}^2+\eta({\bf n} \cdot 
{\bf  p})^2\right ] \frac{1}{r}+\frac{1}{2r^2} \quad ,
\end{eqnarray}
\begin{eqnarray}
H_{2PN}= &\frac{1}{16}& \left (1-5\eta+5\eta^2\right ) \left ({\bf p}^2\right )^3
+\frac{1}{8}\left [\left (5-20\eta-3\eta^2\right)\left ({\bf p}^2\right )^2 
- 2\eta^2({\bf n} \cdot {\bf p})^2{\bf p}^2
-3\eta^2({\bf n} \cdot {\bf p})^4\right ] \frac{1}{r}\nonumber \\
&+&\frac{1}{2}\left [\left (5+8\eta\right ){\bf p}^2
+3\eta({\bf n} \cdot {\bf p})^2\right ] \frac{1}{r^2}
-\frac{1}{4}\left( 1+3\eta \right)\frac{1}{r^3} \ \ .
\end{eqnarray}

The spin-orbit Hamiltonian is
\begin{equation}
H_{SO}=\frac{\bl\cdot {\bs}_{eff}}{r^3}
\end{equation}
with the reduced angular momentum ${\bf L}={\bf r}\times {\bf p}$
and
\begin{equation}
\bs_{eff}=
\left (2+\frac{3m_2}{2m_1}\right ){\bs}_1+
\left (2+\frac{3m_1}{2m_2}\right ){\bs}_2 \ \ .
\end{equation}
Spin-spin coupling terms are not included.
The spins $\bs_i$ are measured in units of $M^2$. To be clear,
$\bs_i={\bf A}_i (m_i/M)^2$ with physical values of the amplitude
${\bf A}_i\le 1$.

The equations of motion are given by
\begin{equation}
\dot {\br}=\frac{\partial H}{\partial {\bf p}} \quad ,\quad
\dot{\bf p}=-\frac{\partial{H}}{\partial {\br}}
\end{equation}
and the evolution equation for the spins and the angular momentum can
be found from the Poisson brackets:
\begin{eqnarray}
\dot {\bs}_1 &=& \{{\bs}_1,H\}=
\left (2+\frac{3m_2}{2m_1}\right )\frac{{\bf L}\times \bs_1}{r^3}\\
\dot {\bs}_2 &=&\{{\bs}_2,H\}  =\left (2+\frac{3m_1}{2m_2}\right
)\frac{{\bf L}\times \bs_2}{r^3}\\
\dot{\bf L} &=& \left \{{\bf L},H\right \}=\frac{{\bs_{eff}}\times{\bf L}}{r^3}
\ \ .
\end{eqnarray}

\begin{figure}
\centerline{\psfig{file=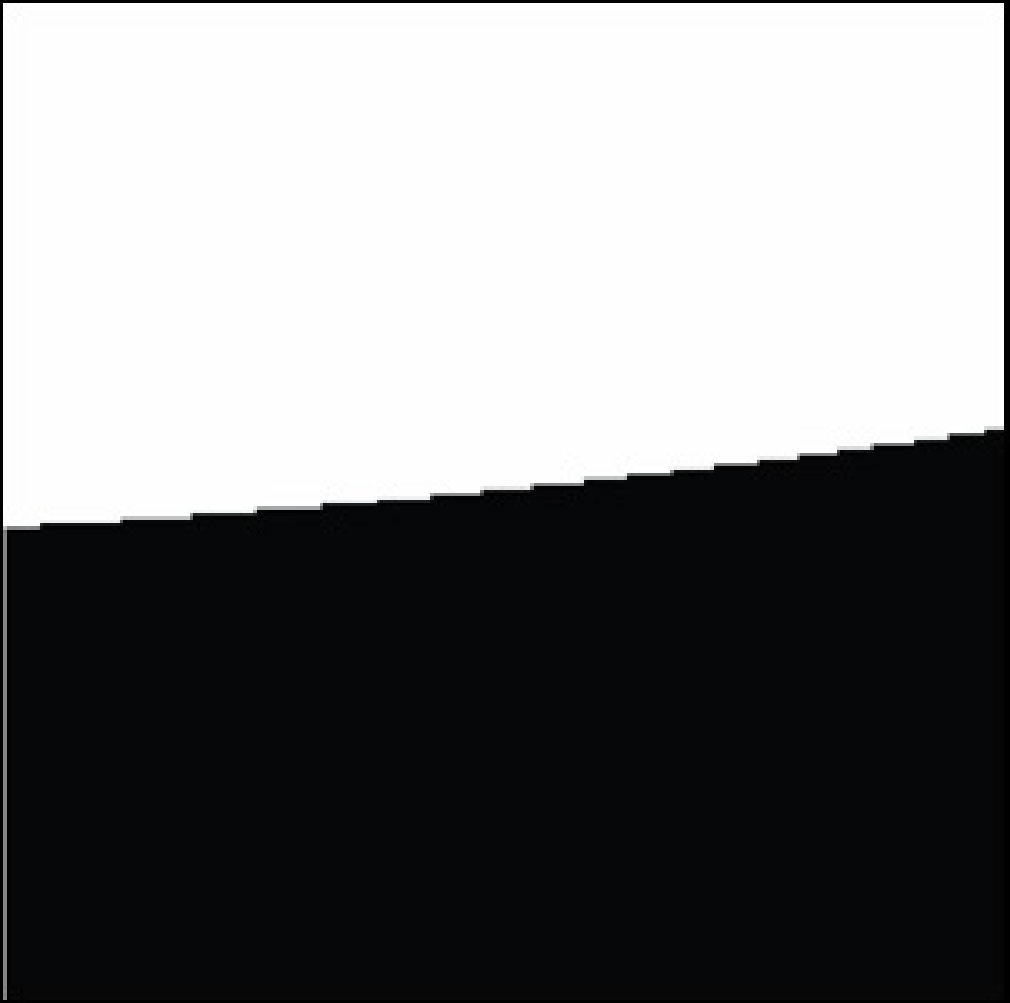,width=2.5in}}
\caption{Basin boundaries in the 2PN-Hamiltonian system with
  spin-orbit couplings. 
The pair has mass ratio $m_2/m_1=1/3$. The heavier black hole 
is maximally spinning 
($S_1= m_1^2$) with an initial angle with respect to the $\hat z$-axis
of $95^o$ 
while the lighter companion is not spinning ($ S_2=0$).
The
initial center of mass separation in ADM coordinates is
$r_i/M=5$. The orbital initial conditions vary along the x-axis from 
$0.02\le p_{r}\le 0.05$ and along the y-axis from $ 0.1195 \le
p_{\phi} \le 0.1200 $. 
$200\times 200$ orbits are shown. Initial conditions that are color-coded
white correspond to stable orbits and those color-coded black
correspond to merging pairs. The basin boundary is clearly smooth
and not fractal. Therefore, there is no evidence of chaos. 
\label{fbbh}}  \end{figure}

A parametric solution for the equations of motion including spin-orbit
  coupling (but not  
  including spin-spin coupling) has been found for one body
  spinning \cite{gk}. Even without a solution, a count of
  conserved quantites shows that motion must lie on a torus. 
Therefore this one case is integrable and should show no
  evidence of chaos. The fractal basin boundary method \cite{cornish}
  confirms that there is no chaos when only one body spins. Figure
  \ref{fbbh} shows the smooth basin between outcomes for such a black
  hole pair. The pair is evolved using the 2PN Hamiltonian including
  spin-orbit couplings for 40,000 different initial conditions. The
  orbits differ in initial $p_r$ and initial $p_\phi$. If the pair
  merge, the initial condition is color coded black. If the pair
  execute more than 50 windings, the initial condition is color coded
  white. The boundary between initial basins - merger and stability -
  is smooth. There is no evidence of chaos in this slice through phase
  space nor any of the others surveyed. Chaos manifests as an extreme
  sensitivity to initial conditions and a mixing of trajectories. A
  smooth boundary shows no such sensitivity or mixing of orbits; that
  is, stable orbits remain on one side of the clean boundary and do
  not mix with unstable orbits, which remain on the other side of the
  smooth boundary.

The exact location and shape
  of the boundary can depend on the exit criteria, but the criteria
  cannot turn a smooth basin into a fractal. For instance, the
  criteria used are that merger occurs when the coordinate $r\le 1$
  and stability corresponds to more than 50 windings. It is worth
  noting that the basin boundaries remain smooth when the
  3PN-Hamiltonian is used as well.

It should be emphasized that while the presence of fractal basin
boundaries provides
an unambiguous signal of chaos, the {\it absence} of fractals at the basin
boundary does not prove the system is integrable. We only show
the smooth basins here to demonstrate consistency between the two
different approaches of Ref.\ \cite{gk} and
\cite{{me},{melong},{menjc}}. 
As no fractals were found despite many scans of
various regions of phase space, the basin boundaries are consistent with
the system being integrable, although it does not provide a proof of
integrability. 

\begin{figure}[ht]
\vspace{65mm}
\includegraphics{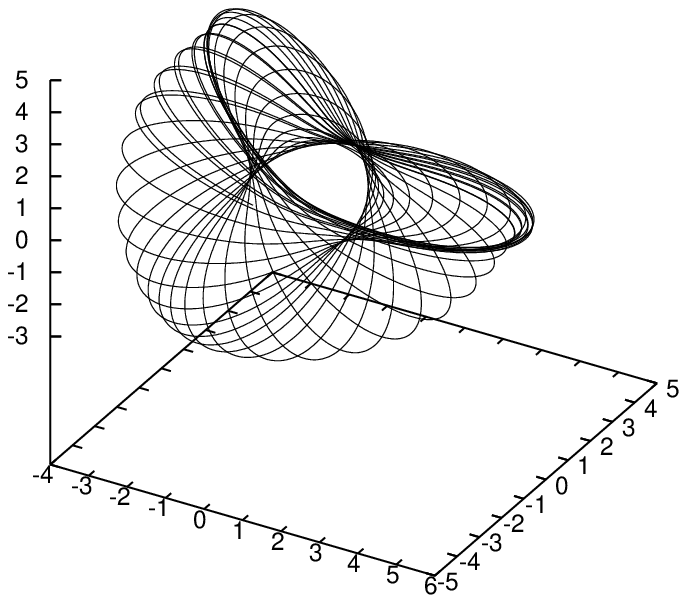}
\includegraphics{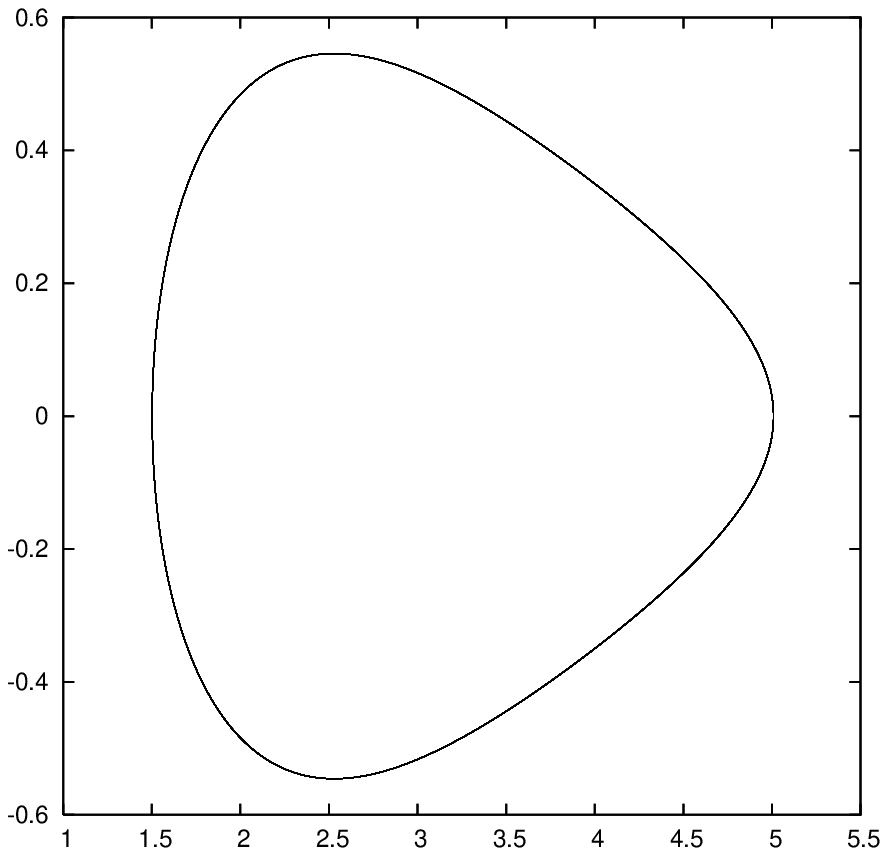}
\caption{Left: An orbit taken from the stable basin of 
fig.\ \ref{fbbh}.
The initial conditions for the orbit are
$p_{r}= 0.03$ and $p_{\phi} = 0.20645 $. 
Right: A projection of the phase space motion onto the $(r,p_r)$
plane. The topology of the projection confirms that 
the motion clearly lies on a torus and therefore is not chaotic.
\label{hamorb}}  
\end{figure} 

An orbit drawn from near the smooth basin boundary of fig.\ \ref{fbbh}
is shown in fig.\ \ref{hamorb}. A Poincar\'e surface of
section could be taken to verify that the motion is confined to a torus 
in phase space. However this isn't necessary since even a projection
of 
the full phase space motion onto the $(r,p_r)$ plane lies on a line
with the topology of a torus. That projection is also shown in fig.\
\ref{hamorb}. As must be the case, both the collective behavior of
orbits and the individual behavior of orbits is regular and not
chaotic.

\section{The PN Lagrangian formulation}

There is chaos
in the
{\it approximately} equivalent Lagrangian formulation of the
dynamics
when only one body spins 
-- for a very narrow range of
parameters \cite{melong}. The chaos appears in some sense
at higher than second order in the PN parameters. 

The equations of motion in harmonic coordinates
\cite{{kidder},{kww}} do not 
correspond to a 
conventional Lagrangian \cite{{damourrev},{damourderuelle}} but 
can be derived from a
generalized Lagrangian that depends on the coordinates, the
velocities, as well as a relative acceleration.
In the following, the variables ($\br$,$\bp$) will continue to refer
to ADM coordinates and their conjugate momenta
while harmonic coordinates and their velocities are denoted
$(\bx,\bv)$ with $\bv=\dot\bx$. 

The generalized
Lagrangian can be expressed in terms of the ADM Hamiltonian. 
In Ref.\ \cite{DJS}, the map 
between the ADM coordinates and the harmonic
coordinates is given explicitly.
Notice that the derivative on ${\br}$ in the ADM Lagrangian will
generate higher derivative acceleration (${\bf \ddot x}$) terms 
when re-expressed in harmonic coordinates:
\begin{equation}
L(\bx,{\bf \dot x},{\bf \ddot x})=
\bp[\bx,{\bf \dot x}]\cdot {\bf \dot r}[\bx,{\bf \dot
    x},{\bf \ddot x}]
-H(\br[\bx,{\bf \dot x}],\bp[\bx,{\bf \dot x}])
\label{lharm} .
\end{equation}
Ref.\ \cite{DJS} presents
the expression for
the harmonic Lagrangian derived from eqn.\ (\ref{lharm}) explicitly.
The Euler-Lagrange equations are 
\begin{equation}
\frac{d}{dt}\left ( \frac{\partial L}{\partial {\bf \dot x}}-\frac{d}{dt}\frac{\partial 
  L}{\partial {\bf \ddot x}} \right ) = \frac{\partial
  L}{\partial \bx}
\ \ .
\end{equation} 
For our purposes it is important to recognize that in order to derive
the Lagrangian equations of motion of Ref.\ \cite{{kidder},{kww}}, wherever the
acceleration appears in higher-order terms in the Euler-Lagrange
equation, the lower-order equation of motion is substituted in its
place.
This is an essential point. As a result of this substitution, the
Lagrangian formulation and the Hamiltonian formulation are only 
{\it approximately} related. Higher-order terms that make them
exactly equal are explicitly discarded to render them only
approximately equal. According to the KAM theorem, while most tori
survive and therefore lead to nonchaotic orbits, there is a
complement set related to the size of the perturbation between them
for which tori are often destroyed leading to chaos.

Furthermore, it is worth noting that for similar reasons the
``constants of motion'' are only {\it approximately} conserved in the
Lagrangian system.
This is to be contrasted with the Hamiltonian
system for which no such substitution is required and the
equations of motion that result from Hamilton's equations {\it
  exactly} conserve the constants of motion. Therefore 
there can never be chaos in the
ADM-Hamiltonian formulation when only one body spins
but {\it there can be chaos} in the
harmonic-Lagrangian equations when only one body spins. 
The chaos must be at a higher order
than the order at which the equations are valid since the constants of
motion are only violated at higher orders. This is consistent with the
KAM theorem statement that tori will only be destroyed for a set of
size related to the size of the perturbation.

Put yet another way -- hammering the point to death -- the
ADM-Hamiltonian equations of motion and the harmonic-Lagrangian
equations of motion are equivalent to 2PN order. However, from a
dynamical systems theory perspective, one can take these two sets of
equations and ask if they are identical, which would require that they
are identical to all orders. Since they are not identical to all
orders, they can show different features and they do. The different
features must be higher than 2PN order, and they are. So a numerical
integration of the two sets of equations of motion will show higher
order differences.

\begin{figure}
\centerline{\psfig{file=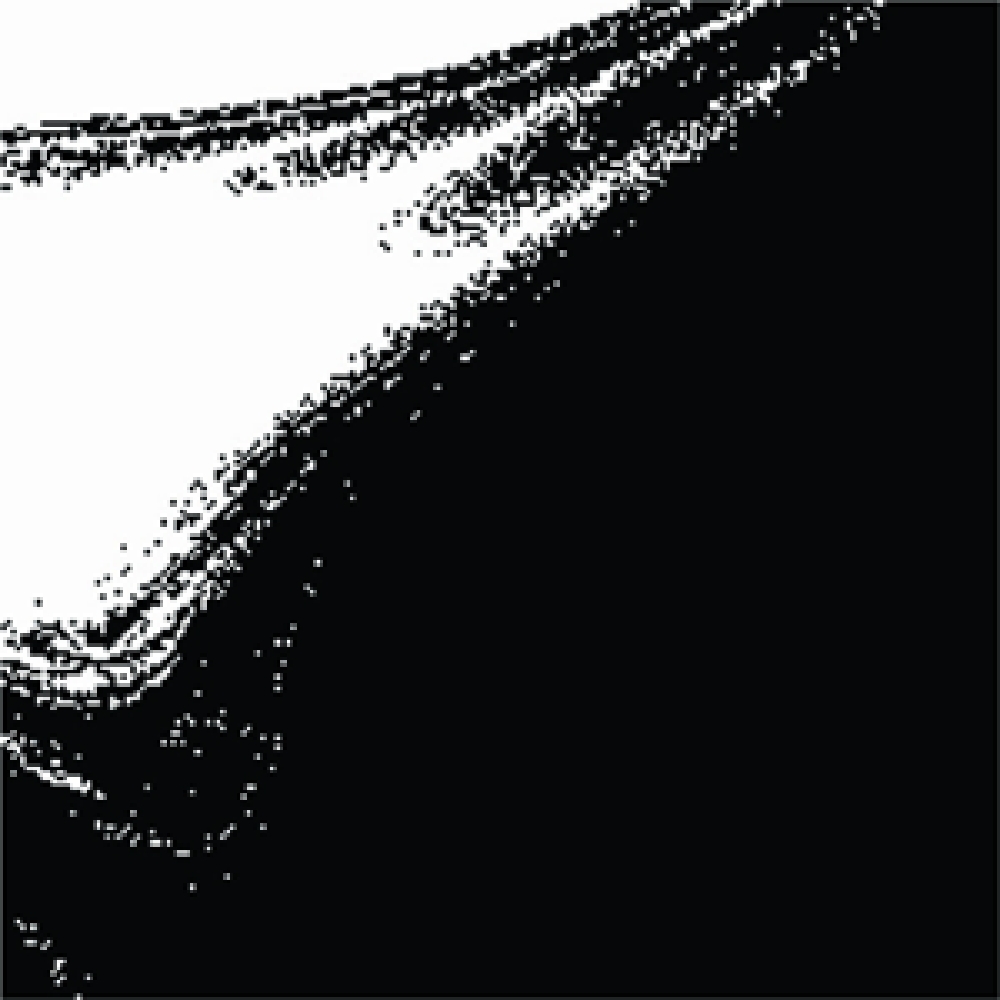,width=2.5in}\quad\quad\psfig{file=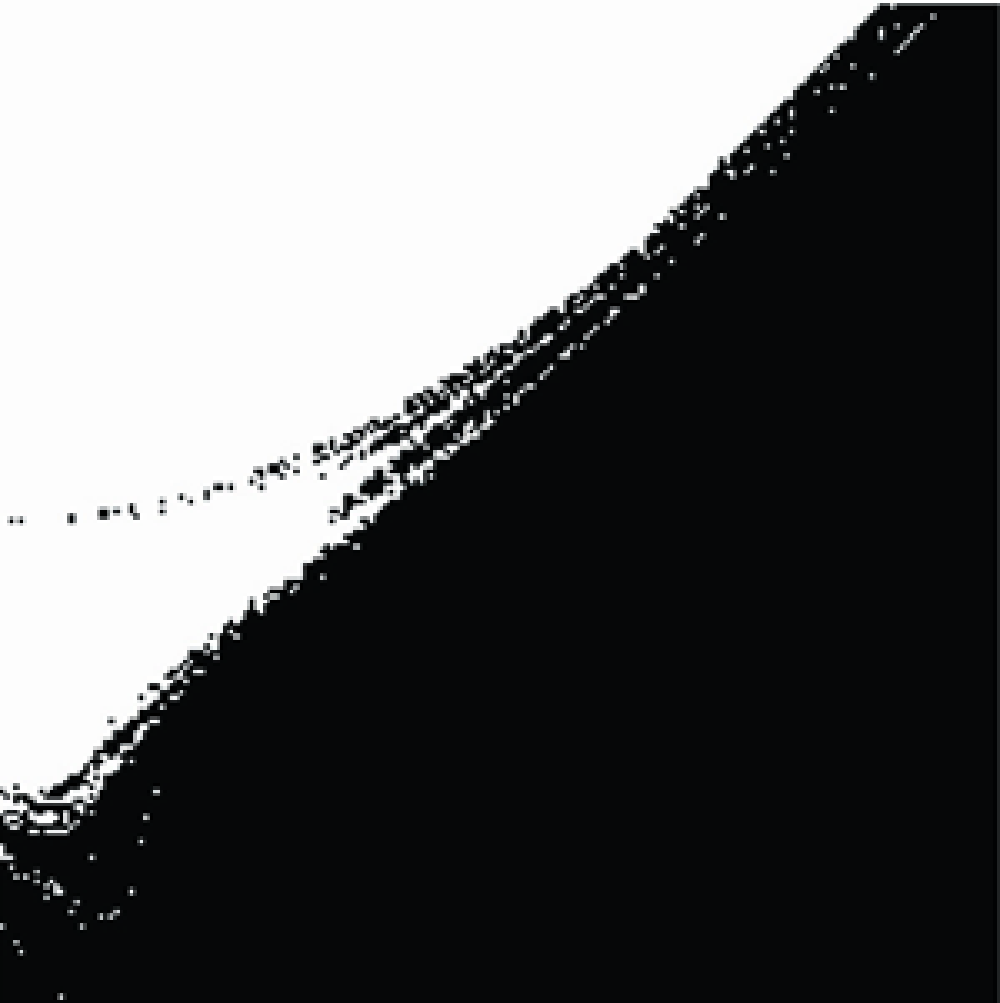,width=2.5in}}
\caption{
The pair has mass ratio $m_2/m_1=1/3$. 
The initial center of mass separation in harmonic coordinates is $5M$.
Using the notation $\bx=(x^{(1)},x^{(2)},x^{(3)})$,
the orbital initial conditions vary along the x-axis from 
$0\le \dot x^{(1)}\le 0.035$ and along the y-axis from $ 0.425 \le \dot
x^{(2)} \le 0.443125 $. 
$200\times 200$ orbits are shown. 
Initial conditions that are color-coded
white correspond to stable orbits and those color-coded black
correspond to merging pairs. The basin boundary is fractal indicating
at least a thin region of chaos. 
Left: A reproduction of Fig.\ 4 from Ref.\ \cite{melong}.
The heavier black hole 
is maximally spinning 
($S_1= m_1^2$) with an initial angle with respect to the $\hat z$-axis
of $95^o$ 
while the lighter companion is not spinning ($ S_2=0$).
Right: The heavier black hole 
is not spinning ($S_1=0$) while the lighter object is maximally spinning 
($S_2= m_2^2$) with an initial angle with respect to the $\hat z$-axis
of $95^o$ .
\label{fbbkww}}  \end{figure} 

The resultant Lagrangian equations of motion in $\bx,\bv$ with spins added are 
(with $x=\sqrt{\bx\cdot\bx}$ the harmonic radial coordinate)
\cite{{kidder},{kww}} 
\begin{eqnarray}
\label{keom1}
\dot \bx &=& \bv \\
\ddot \bx &=& {\bf a}_N+{\bf a}_{1PN}+{\bf a}_{2PN}+{\bf a}_{SO}
\end{eqnarray}
and
\begin{eqnarray}
\label{keom2}
{\bf a}_N &=&-\frac{\bn}{x^2}\nonumber \\
{\bf a}_{1PN} &=&-\frac{\bn}{x^2}\left
\{(1+3\eta)\bv^2-2(2+\eta)\frac{1}{x}-\frac{3}{2}\eta\dot x^2\right
\}-\frac{\bv}{x^2}2(2-\eta)\dot x\nonumber \\
{\bf a}_{2PN}&=&-\frac{\bn}{x^2}\left\{\frac{3}{4}(12+29\eta)\frac{1}{x^2}+\eta(3-4\eta)(\bv^2)^2+\frac{15}{8}\eta(1-3\eta)\dot
x^4\nonumber\right. \\
 & {}&-\left.\frac{3}{2}\eta(3-4\eta)\bv^2\dot
x^2-\frac{1}{2}\eta(12-4\eta)\frac{\bv^2}{x}-(2+25\eta+2\eta^2)\frac{\dot
  x^2}{x}\right \}\nonumber \\
&-&\frac{\bv}{x^2}\left \{-\frac{\dot x}{2}\left
[\eta(15+4\eta)\bv^2-(4+41\eta+8\eta^2)\frac{1}{x}-3\eta(3+2\eta)\dot
  x^2\right ]\right \} \quad .
\end{eqnarray}
The spin-orbit contribution to the acceleration is
\begin{equation}
\nonumber
{\bf a}_{SO}=\frac{1}{x^2}\left \{6\bn\left [(\bn\times\bv)\cdot\left
  (2\bs+\frac{\delta M}{M}\bD \right )\right ]-\left [\bv\times
  \left (7\bs+3\frac{\delta M}{M}\bD\right )\right ]
+3\dot x\left [\bn \times \left (3\bs+\frac{\delta M}{M}\bD \right
    )\right ]\right \} \nonumber 
\end{equation}
with $\bD=M\left(\bs_2/m_2-\bs_1/m1\right )$, $\eta=\mu/M$ and $\delta
M=m_1-m_2$.
There are also spin-spin terms but they vanish in the event that one
of the black holes is spinless and so are omitted here in the
comparison.
Finally,
the spins precess according to
\begin{eqnarray}
\dot \bs_1&=&\frac{(\bx\times\mu\bv)\times \bs_1}{x^3}\left
(2+\frac{3m_2}{2m_1}\right ) \nonumber \\
\dot \bs_2&=&\frac{(\bx\times\mu\bv)\times \bs_2}{x^3}\left
(2+\frac{3m_1}{2m_2}\right ) \ \ .
\label{kwwss}
\end{eqnarray}
Equations (\ref{keom1})-(\ref{kwwss}) constitute the Lagrangian
dynamical system.

Fig.\ \ref{fbbkww} shows fractal boundaries at the basins of stability
and merger in a slice through phase space. The fractal boundaries
prove that there is chaotic scattering in these region of phase space
for the Lagrangian approximation
\cite{ott}. Two basins are shown for black holes with a mass ratio
$m_2/m_1=1/3$. For the basin boundaries shown on the left of
fig.\ \ref{fbbkww}, the heavier
black hole is maximally 
spinning while the lighter companion has no spin. For the basin
boundaries on the right of the figure, the
lighter black hole is maximally spinning while the heavier black hole
has no spin. In both cases, the boundaries show extreme sensitivity to
initial conditions and a mixing of orbits. 

To be clear, this is the same physical
system (although it is not the exact same slice through phase space)
as that shown 
in fig.\ \ref{fbbh} for the Hamiltonian approximation. But unlike
the Hamiltonian approximation, which always had smooth boundaries, the
Lagrangian approximation shows fractal boundaries. 
It is also essential to
note that these boundaries are extremely thin. For the slice through
phase space in fig.\ \ref{fbbkww},
1PN corrections are roughly $v^2<0.2$, 2PN corrections are
roughly $v^4<0.04$, and
so corrections higher than the order at which the approximation is
valid appear around $v^6<0.008$.
The sensitivity to
initial conditions seen in the boundaries only appears for differences in
initial velocity of $\ll 0.005$, which indicates that 3PN corrections
are at least this large. The 2PN approximation is being pushed beyond
its limits. Again, this speaks to the consistency of the two
approximations at the 2PN level. The conclusion can only be that there is chaos
in the 2PN Lagrangian-approximation when only one body spins but it is
difficult to draw physical conclusions since it only appears at such
high 
decimal places.

\begin{figure}[ht]
\vspace{65mm}
\includegraphics{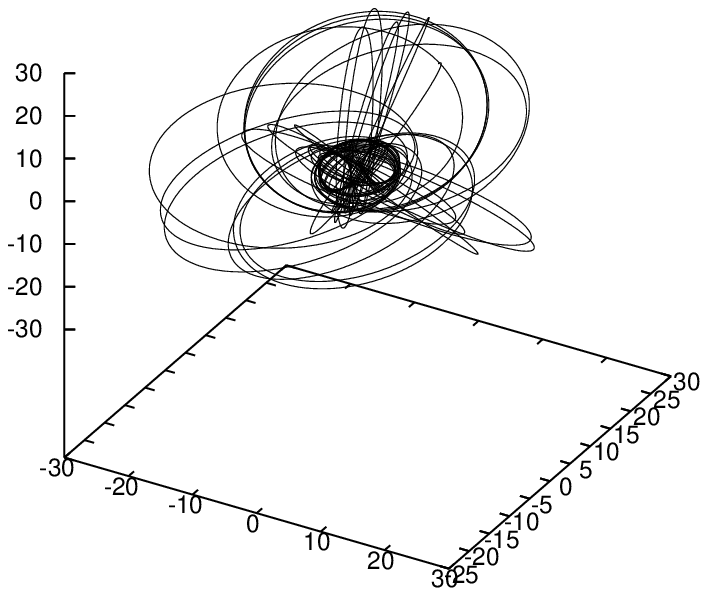}
\includegraphics{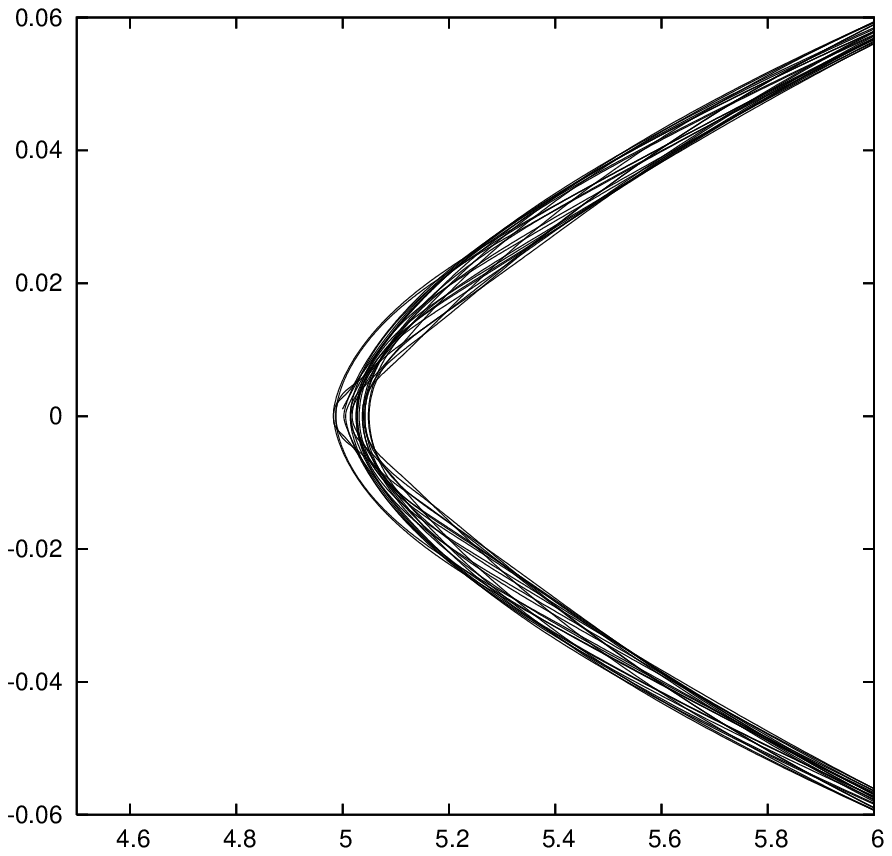}
\caption{Left: An orbit taken from the stable basin of 
fig.\ \ref{fbbkww}.
The initial conditions for the orbit are
$\dot x^{(1)}= 0.00105$ and $\dot x^{(2)} = 0.43074 $. 
Right: A detail of the projection of the motion in phase space onto
the
radial
$(x,\dot x)$ plane, where again $x=\sqrt{\bx\cdot\bx}$.
\label{kwworb}}  
\end{figure} 

An orbit drawn from the stable basin of fig.\ \ref{fbbkww} is shown in
fig.\ \ref{kwworb}. Also shown is a detail of the projection of the
motion onto the radial $(x,\dot x)$ plane, where again
$x=\sqrt{\bx\cdot\bx}$. The projection does not 
lie on a line with the topology of a torus, indicating that the motion
may not be regular.
Now being careful, it might be the case that the spread off
of a torus shown in this detail is an artifact of projecting the
multi-dimensional coordinate space down to the radial $(x,\dot x)$
plane. Consequently,  
this does not prove that the motion is lifted off of a torus, it only
suggests that it has diffused off of a torus. It is therefore not
proof of chaos, only a suggestion of chaotic motion. A proper
Poincar\'e surface of section in a properly defined phase space would
need to be taken to prove the orbit 
is chaotic. Instead however, we confirmed
that this orbit does have a positive Lyapunov
exponent \cite{{menjc_comment},{menjc}}, which proves that this orbit is
indeed chaotic.  

Not all of the orbits drawn from the stable basin will be
chaotic. Some will remain on tori in phase space and will have zero Lyapunov
exponent. Those that do show chaos seem to lie very close to the
already thin fractal basin boundary. This again confirms the
expectation that those tori that are destroyed occupy a very thin
region of phase space corresponding to perturbations to the
Hamiltonian system that are higher than second order.

\section{summary}

In the previous section it was shown that the same methods applied to
the 2PN-Hamiltonian formulation and the 2PN-Lagrangian formulation
found no chaos in the former and chaos in the latter. There is no
conflict between these results as the Hamiltonian approach and the
Lagrangian approach are only approximately related. One of the very
underpinnings of the development of chaos theory is the realization
that a regular Hamiltonian system can become chaotic under small
perturbation. The Lagrangian formulation can be viewed as a small
perturbation to the Hamiltonian system and so the emergence of chaos
is permitted -- if subtle.

It could be argued that the Hamiltonian approach is more appealing
analytically since the constants of motion are exactly conserved. This
is always an advantage, particularly in numerical studies in which
the constants can be tracked and their constancy continually checked.
The derivation of the equations of motion is also cleaner and more
direct. 
Regardless,
we have not resolved the physical question: Is 
there chaos in the orbits of comparable mass binaries when only one
spins? 
It is highly likely that when spin-spin terms are included, the
Hamiltonian approximation will also show chaos even for
the special case of one body spinning -- although the effect may continue to 
appear at orders higher than the approximation can be trusted.
All we can be sure of at this point is that we have
another reflection of the poor
convergence of 
the PN expansion to the full relativistic system. In our hopes to
provide gravitational wave templates for the gravitational wave
observatories, we are tempted to push these approximations into regimes
where they are not faring well \cite{hughes}. In an ideal world, we
would have an excellent approximation that remained valid at small
separations, near the innermost stable circular orbit, and in the most
highly nonlinear regime. That we do not have. In the interim, all we
can do is treat the approximations as dynamical systmes and see what
emerges. And quite fascinatingly, what has emerged is two different
claims on the chaotic behavior of comparable mass binaries.

Just to be clear, there is inarguably chaos at physically accessible values of
the spins when {\it both objects spin}. This was shown in the initial papers 
\cite{{me},{melong},{menjc}} and was verified in the Hamiltonian
formulation in Ref.\ \cite{HB}. The authors of Ref.\ \cite{HB} were able to
investigate a wide range of parameters to conclude that chaos did not
appear to have a huge impact on gravitational waves in the LIGO
bandwidth. Similarly, we have shown that the damping effects of
dissipation when the 
radiation reaction is included \cite{{menjc_comment},{menjc},{menjc2}}
can squelch chaos.  
One can imagine that these
conclusion will remain more or less intact even at higher order. But we
cannot know for certain. One can also imagine that 
the chaos will worsen as our currently poor approximations to the
full relativistic problem improve. My projection is that the chaos
will 
worsen a bit but will not plague the ground-based
gravitational wave observatories in a
detrimental way. Regardless, chaos
will continue to be important primarily at very late stages of
inspiral, such as the transition from
inspiral to plunge, and cannot be disregarded in our theoretical
attempts to understand the dynamics of black hole pairs.

\vskip 25truept

\centerline{\bf Acknowledgements}

I am grateful to the following people for their valuable input:
Adam Brown, Neil Cornish,Becky Grossman,Gabe Perez-Giz, and Julia Sandell.


\begin{thebibliography}{99}
\bibitem{note} {Incidentally, in Ref.\ \cite{gk} a
  distinction is 
  drawn between derminism and chaos. However, it should be noted that chaotic
  systems are also truly deterministic. There is cause and there is
  effect. There are equations of motion that lead to determined output
given any specific input. Chaos refers to nonintegrability -- the
nonexistence of a smooth analytic function that interpolates between
orbits -- and does not refer to a system being non-deterministic.}
\end{thebibliography}
\end{document}